\documentclass[pra,showpacs,superscriptaddress,amsfonts,amsmath,floatfix]{revtex4}

\usepackage{graphicx}
\usepackage{color}

\begin{document}

\title{Loschmidt echo in one-dimensional interacting Bose gases}

\author{K. Lelas}\email{klelas@fesb.hr}
\affiliation{Faculty of Electrical Engineering Mechanical
Engineering and Naval Architecture, University of Split, Rudjera
Bo\v{s}kovi\'{c}a BB, 21000 Split, Croatia}
\author{T. \v Seva}\email{tseva@phy.hr}
\affiliation{Department of Physics, University of Zagreb,
Bijeni\v{c}ka c. 32, 10000 Zagreb, Croatia}
\author{H.~Buljan}\email{hbuljan@phy.hr}
\affiliation{Department of Physics, University of Zagreb,
Bijeni\v{c}ka c. 32, 10000 Zagreb, Croatia}

\date{\today}

\begin{abstract}
We explore Loschmidt echo in two regimes of one-dimensional (1D)
interacting Bose gases: the strongly interacting Tonks-Girardeau
(TG) regime, and the weakly-interacting mean-field regime. We find that the
Loschmidt echo of a TG gas decays as a Gaussian when small perturbations 
are added to the Hamiltonian (the exponent is proportional to the number of 
particles and the magnitude of a small perturbation squared). 
In the mean-field regime the Loschmidt echo decays faster for 
larger interparticle interactions (nonlinearity), and it shows richer 
behavior than the TG Loschmidt echo dynamics, with oscillations 
superimposed on the overall decay. 
\end{abstract}

 \maketitle

\section{Introduction}

The understanding of why an isolated (interacting many-body) system,
which is initially say far from equilibrium, in many cases macroscopically 
undergoes irreversible evolution towards an equilibrium state, despite the 
fact that the microscopic laws are reversible,
has intrigued scientists ever since the first disputes between
Boltzmann and Loschmidt on this topic \cite{Loschmidt, Boltzman}. 
In principle, if the time was reversed at a given instance, the system
would evolve back into the initial state. However, such a reversal
is for all practical purposes impossible due to high sensitivity to 
small errors and interaction of the system with the 
environment. The quantity that measures sensitivity of quantum motion 
to perturbations is called {\em Loschmidt echo} or {\em fidelity} 
\cite{Peres,Rodolfo,Jacquod,Cerruti,Prosen}
(for a review see e.g. \cite{Gorin}). Fidelity tells us what is
the probability that system will end up in the initial state after
forward evolution for time $t$, followed by the slightly imperfect time
reversed evolution for the same time $t$. In quantum mechanics time
evolution from an initial state $\psi_0$ is given by the unitary operator
$\hat{U}(t)=\exp(-i\hat{H}t/\hbar)$ via 
$\psi(t)=\exp(-i\hat{H}t/\hbar)\psi_0$, and the echo dynamics can be
formally written as
\begin{equation}
F(t)=|\langle\psi_0|\exp(i\hat{H'}t/\hbar)\exp(-i\hat{H}t/\hbar)|\psi_0\rangle|^2.
\end{equation}
Here $\hat{H}$ is the Hamiltonian of the unperturbed system, and 
$\hat{H'}=\hat{H}+\hat{V_{\varepsilon}}$ is the slightly perturbed 
Hamiltonian \cite{Gorin}. One can think about this quantity as 
measuring the stability of quantum motion \cite{Peres}, i.e., it tells us the 
overlap of the two states $\psi(t)$ and $\psi'(t)$, the former is evolved 
forward in time by $\hat{H}$, and the latter by $\hat{H'}$:
\begin{equation}
F(t)=|\langle\psi'(t)|\psi(t)\rangle|^2. \label{fidelity}
\end{equation}
In quantum systems, depending on the strength of the perturbation 
and the properties of the nonperturbed Hamiltonian, 
three different decay regimes of Loschmidt echo are usually identified: 
the Gaussian perturbative regime \cite{Jacquod, Cerruti}, 
the exponential Fermi golden rule regime \cite{Jacquod, Cerruti, Prosen, Rodolfo}, 
and the Lyapunov regime \cite{Rodolfo}.

Motivated by the recent progress in experiments and theory on 
ultracold atomic gases \cite{Bloch2008}, where the influence of the 
environment can be made very small, and the strength of the 
atom-atom interactions can be tuned \cite{Bloch2008}, we are motivated 
to investigate Loschmidt echo dynamics in those systems. 
In particular, we focus on one-dimensional (1D) Bose gases 
which were experimentally realized \cite{OneD} even in the strongly 
correlated regime of Tonks-Girardeau (TG) bosons \cite{TG2004, Kinoshita2006}, 
both in \cite{TG2004} and out of equilibrium \cite{Kinoshita2006}. 
The realization of the TG gas in atomic waveguides was proposed by Olshanii \cite{Olshanii}.
The 1D atomic gases can be described with the Lieb-Liniger model \cite{Lieb1963}, 
which for weak interactions is well-described by the Nonlinear Schr\"{o}dinger 
equation (Gross-Pitaevskii theory) \cite{Bloch2008}, whereas for sufficiently 
strong interactions one enters the Tonks-Girardeau regime, 
where exact solutions can be found via Fermi-Bose mapping \cite{Girardeau1960}. 
This method was used to study out-of-equilibrium dynamics in the 
strongly correlated regime (e.g., see \cite{Girardeau2000, Rigol2005, Minguzzi2005, 
Rigol2006, delCampo2006, Pezer2007}). 
The Loschmidt echo was within the mean-field Gross-Pitaevskii theory addressed in  
Refs. \cite{Manfredi2008, Martin2008}. 
We would also like to point out at a study of orthogonality 
catastrophe (and the relation to Loschmidt echo) 
in an ultracold Fermi gas coupled to a single cubit 
\cite{Goold2011}.

Here we demonstrate, with exact numerical calculation, that for small
random stationary perturbation, the Loschmidt echo for a TG gas decays 
as a Gaussian with decay constant proportional to the number of particles and 
the square of the amplitude of the perturbation. 
We analytically derive the Gaussian behavior of TG fidelity within 
approximation presented by Peres \cite{Peres}. 
In the mean-field regime the Loschmidt echo decays faster for 
larger interparticle interactions (nonlinearity), and it shows richer 
behavior than the TG Loschmidt echo dynamics, with oscillations 
superimposed on the overall decay.

\section{The physical system and the corresponding model}

Consider a gas of $N$ identical bosons in a one-dimensional (1D)
space, which interact via pointlike interactions, described by the
Hamiltonian
\begin{equation}
H=\sum_{i=1}^{N} \left[-\frac{\hbar^2}{2m}\frac{\partial^2}
{\partial X_{i}^2}+U(X_{i}) \right]+g_{1D}\sum_{1\leq i<j \leq N}
\delta(X_i-X_j). \label{Hamiltonian}
\end{equation}
Such a system can be realized with ultracold bosonic atoms trapped
in effectively 1D atomic waveguides \cite{OneD,TG2004,Kinoshita2006}, 
where $U(X)$ is the axial trapping potential, and $g_{1D}=2\hbar^2
a_{3D}[ma_{\bot}^2(1-Ca_{3D}/\sqrt{2}a_\bot)]^{-1}$ is the
effective 1D coupling strength; $a_{3D}$ stands for the
three-dimensional s-wave scattering length,
$a_\bot=\sqrt{\hbar/m\omega_\bot}$ is the transverse width of the
trap, and $C=1.4603$ \cite{Olshanii}. By varying
$\omega_\bot$ the system can be tuned from the mean field regime
described by the Gross-Pitaevskii equation, up to the
strongly-interacting Tonks-Girardeau regime
($g_{1D}\rightarrow\infty$). In equilibrium, different regimes of
these 1D gases are usually characterized by a dimensionless
parameter $\gamma =mg_{1D}/\hbar^2 n_{1D}$, where $n_{1D}$ stands
for the linear atomic density. For $\gamma\ll 1$ the gas is in mean
field regime and for $\gamma\gg 1$ it is in the strongly interacting
regime (we consider repulsive interactions $\gamma>0$) \cite{Lieb1963, Olshanii, 
TG2004, Kinoshita2006}. One can tune $\gamma$ by say changing the transverse 
confinement frequency.
In our calculations, we use Hamiltonian (\ref{Hamiltonian}) in its
dimensionless form
\begin{equation}
H=\sum_{i=1}^{N} \left[-\frac{\partial^2}{\partial x_{i}^2}+V(x_{i})
\right]+ 2c\sum_{i<j}^N \delta(x_i-x_j). \label{HamiltonianDless}
\end{equation}
where $x=X/X_0$ ($X_0$ is the spatial scale which we choose to be
$1\,\mu$m). Here we consider $^{87}$Rb atoms with 
the 3D scattering length $a_{3D}=5.3$~nm \cite{TG2004,Kinoshita2006}. 
Energy is in units of $E_0=\hbar^2/2mX_0^2=3.82\cdot10^{-32}$~J, and time 
is in units of $T_0=2mX_0^2/\hbar=2.8$~ms.  The
dimensionless axial potential is $V(x)=U(X)/E_0$, and the
interactions strength parameter is $2c=g_{1D}/X_0 E_0$.

\section{Loschmidt echo of a Tonks-Girardeau gas}

For the Tonks-Girardeau gas, the interaction strength is infinite
$c\rightarrow \infty$, that is, the bosons are "impenetrable"
\cite{Girardeau1960}. Consequently, an exact (static and time-dependent)
solution of this model can be written via Girardeau's Fermi-Bose
mapping \cite{Girardeau1960, Girardeau2000}
\begin{equation}
\psi_B(x_1,\ldots,x_N,t)=\prod_{1\leq i<j\leq N}
\mbox{sgn}(x_i-x_j)\psi_F(x_1,\ldots,x_N,t), \label{FBmapping}
\end{equation}
where $\psi_F$ denotes a wave function describing $N$ noninteracting
spin polarized fermions in the external potential $V(x)$. In our
simulations we consider up to $N=70$ particles. The system is
initially (for times $t\leq 0$) in the ground state of a container
like potential,
\begin{equation}
V_{L}(x)=V_0 \{ 1 + \tanh[V_s(x-L/2)]/2-\tanh[V_s(x+L/2)]/2 \}, \label{pot}
\end{equation}
where $V_0=500$, $V_s=4$, and $L=15$ (corresponding to $15\,\mu$m).
At $t=0$ we suddenly expand the width of the container to twice its
original width, that is, the potential at times $t>0$ is
$V_{2L}(x)$. In order to calculate the fidelity $F(t)$, we must
evolve the TG gas in the new potential $V_{2L}(x)$, and in the
potential $V_{2L}'(x)=V_{2L}(x)+V_{\varepsilon}(x)$ starting from
identical initial states. Here $V_{\varepsilon}(x)$ is a small noise
potential of amplitude $\varepsilon$. The Loschmidt echo
$F(t)=|\langle\psi_B'(t)|\psi_B(t)\rangle|^2$ is calculated from the
knowledge of the TG many-body states $\psi_B(t)$ and $\psi_B'(t)$
corresponding to the evolution in potentials $V_{2L}(x)$ and
$V_{2L}'(x)$, respectively.

The fermionic wave function $\psi_F$ can in our case be written as a
Slater determinant,
$\psi_F(x_1,\ldots,x_N,t)=\det_{m,n=1}^N[\psi_m(x_n,t)]/\sqrt{N!}$,
where $\psi_m(x,t)$, $m=1,\ldots,N$, satisfy the single-particle
Schr\" odinger equation
\begin{equation}
i \frac{\partial \psi_m(x,t)}{\partial t}=
\left[-\frac{\partial^2}{\partial x^2}+V_{2L}(x) \right]\psi_m(x,t),
\label{singlesch}
\end{equation}
and equivalently for $\psi_m^{'}(x,t)$ which evolve in $V_{2L}'(x)$.
The initial conditions are such that $\psi_m(x,0)=\psi_m^{'}(x,0)$
is the $m$-th single particle eigenstate of the initial container
potential $V_L(x)$. The Loschmidt echo for a Tonks-Girardeau gas can
be written in a form convenient for calculation:
\begin{eqnarray}
|\langle\psi_B'(t)|\psi_B(t)\rangle|^2 & = & |\frac{1}{N!} \int dx_1
\cdots dx_N \sum_{\sigma_1}(-)^{\sigma_1} \prod_{i=1}^N
\psi^{'*}_{\sigma_1(i)}(x_i,t)
\sum_{\sigma_2}(-)^{\sigma_2} \prod_{j=1}^N \psi_{\sigma_2(j)}(x_j,t)|^2 \nonumber \\
& = & |\frac{1}{N!}
\sum_{\sigma_1}\sum_{\sigma_2}(-)^{\sigma_1}(-)^{\sigma_2}
\prod_{i=1}^N P_{\sigma_1(i)\sigma_2(i)}(t)|^2 \nonumber \\
& = & |\det\mathbf{P}(t)|^2, \label{fidelityTG}
\end{eqnarray}
where $\sigma$ denotes a permutation in $N$ indices, $(-)^{\sigma}$
is its signature, and
\begin{equation}
P_{ij}(t)=\int\psi_i^{'*}(x,t)\psi_j(x,t)dx \label{matrixP}.
\end{equation}
In writing relation (\ref{fidelityTG}) we used a definition of the 
determinant. 
Since at $t=0$ we have $P_{ij}(0)=\delta_{ij}$, that motivates us to define 
the fidelity product
\begin{equation}
F_P(t)=\prod_{i=1}^N P_{ii}(t)P_{ii}^*(t). \label{FP}
\end{equation}
Thus, in calculation of the fidelity product we assume that all off
diagonal elements of the matrix (\ref{matrixP}) are zero i.e
$P_{ij}(t)=0$ for $i\neq j$ for all times. It can be interpreted as
if we evolve the $N$ particles fully independently of each other
(including statistics) starting from the $N$ initial states
$\psi_m(x,0)$, $m=1,\ldots,N$, calculate $N$ different fidelities
for these states, and multiply them to obtain the product fidelity. 
The value $F(t)$ is identical for noninteracting spinless fermions and 
interacting TG bosons; note that Eq. (\ref{fidelityTG}) is {\em identical} 
to the formula used by Goold {\it et al.} in Ref. \cite{Goold2011} studying 
the orthogonality catastrophe for ultracold fermions. 
Thus, $F_P$ and $F$ will distinguish the influence of
antysymmetrization in the case of noninteracting fermions, or TG
interactions and symmetrization in the case of bosons, with dynamics
which does not include neither statistics nor interactions 
into account. We would like to
emphasize that derivation of (\ref{fidelityTG}) and (\ref{matrixP})
does not require that we initiate the dynamics from the ground state of the TG gas
in the initial trap; we could have chosen any excited TG eigenstate 
as initial condition as well.

In order to calculate the fidelity $F(t)$, we must evolve the single
particle states $\psi_j(x,t)$ [$\psi_j^{'}(x,t)$, respectively], in
the potential $V_{2L}(x)$ [$V_{2L}(x)+V_{\varepsilon}(x)$], starting
from the first $N$ single-particle eigenstates of $V_{L}(x)$. The
evolution is performed via standard linear superposition in terms
of the eigenstates $\phi_m(x)$ of the final container potential
$V_{2L}(x)$ (which are calculated numerically):
\begin{equation}
\psi_j(x,t)=\sum_m a_m^j \phi_m(x)\exp(- i E_m t),
\end{equation}
and
\begin{equation}
\psi_j'(x,t)=\sum_m a_m^{j'} \phi_m'(x)\exp(- i E_m' t).
\end{equation}
We numerically calculate the coefficients
$a_{n}^{j'}=\int\phi_{m}^{'*}(x)\psi_j(x,0) dx$ and
$a_{n}^{j}=\int\phi_{m}^{*}(x)\psi_j(x,0) dx$ for $j=1,\ldots,70$,
and $m=1,\ldots,210$ which is sufficient for the parameters we used. 
The noise potential
is constructed as follows: $x$-space is numerically simulated by
using $2048$ equidistant points in the interval $x\in[-30,30]$. From
this array we construct a random array
$V_{\textrm{rand}}(x)=|FT^{-1}[\exp(-k^4/K_{cut}^4)FT[\textrm{rand}(x)]]|$ of the same
length, where $\textrm{rand}(x)$ is a random number between $0$ and $1$,
$FT$ stands for the Fourier transform, $K_{cut}$ is the cut-off wave vector
(set to $K_{cut}=53$) introduced to make the discrete numerical potential 
sufficiently "smooth" from point to point. 
Finally the noise potential is obtained via 
$V_{\varepsilon}(x)=\varepsilon[V_{\textrm{rand}}(x)-\bar{V}_{\textrm{rand}}]$, 
where $\varepsilon$ is the amplitude of the perturbation, and
$\bar{V}_{\textrm{rand}}$ is the mean value of $V_{\textrm{rand}}(x)$.
Such a potential can be constructed optically for 1D Bose gases \cite{Billy2008}.

The fidelity depends on the particle number $N$, the amplitude of
noise $\varepsilon$, but also on the particular realization of
$V_{\varepsilon}(x)$, hence, we calculate all quantities (e.g., the
Loschmidt echo) for 50 different realizations of
$V_{\varepsilon}(x)$, and then perform the average over the noise
ensemble: $\langle F(t) \rangle_{noise}$.

\begin{figure}[!h]
\centering
\includegraphics[scale=0.45]{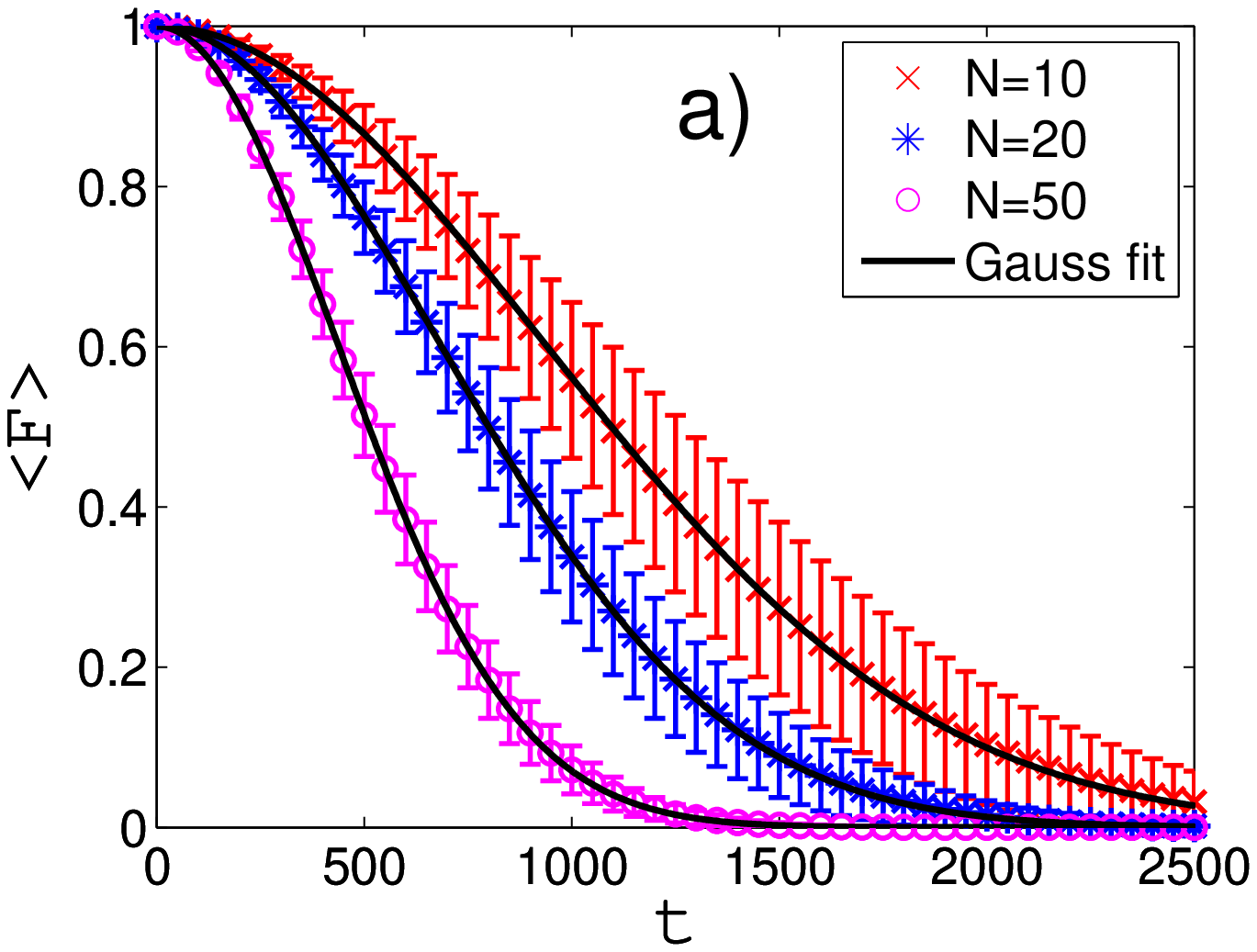}
\includegraphics[scale=0.45]{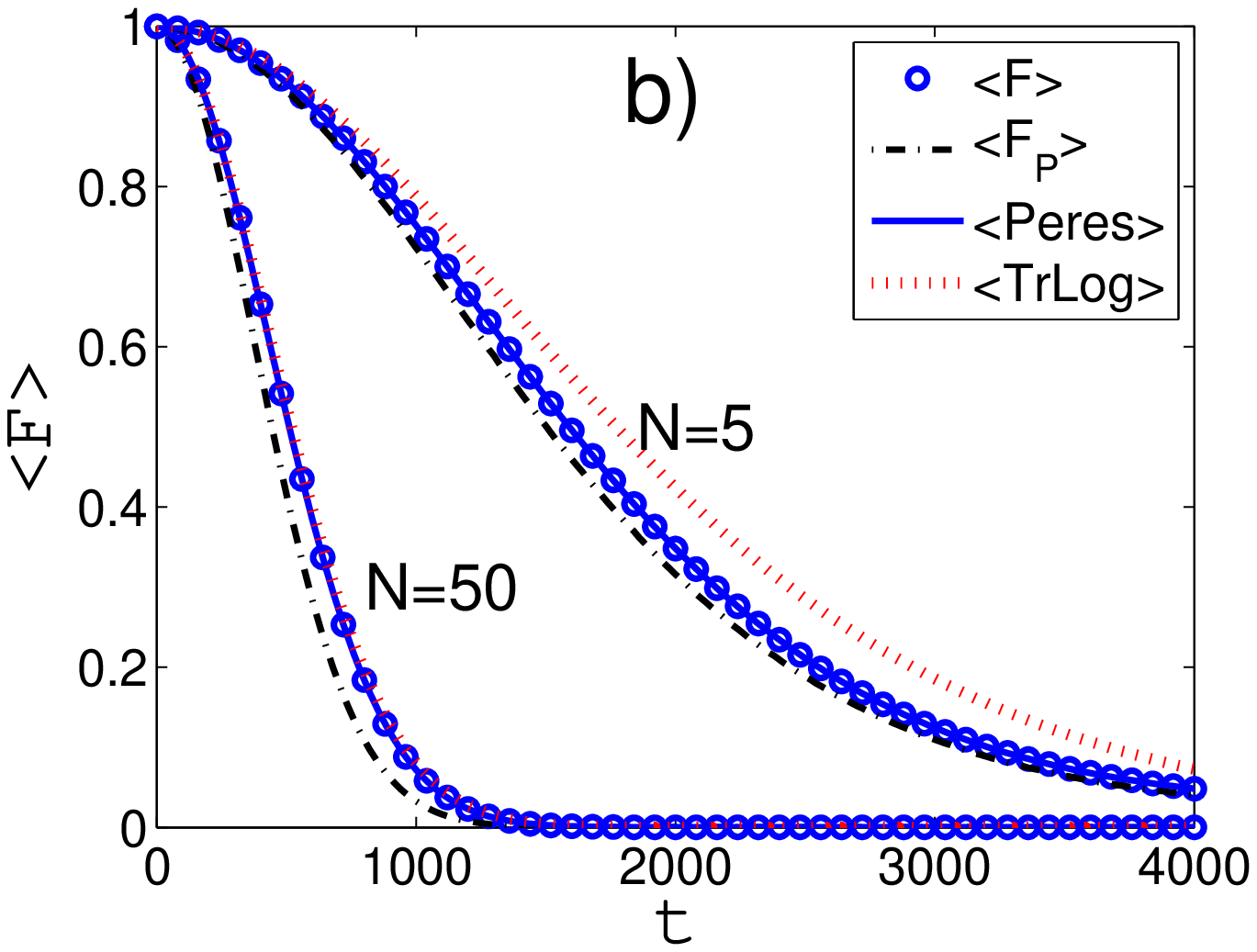}
\caption{(color online) Decay of the Loschmidt echo (fidelity) with
time for $\varepsilon=0.05$. (a) The averaged values $\langle F(t) \rangle_{noise}$ for
$N=10$ (crosses), $N=20$ (asterisks), and $N=50$ (circles). Solid
black lines represent the Gaussian functions fitted to the
numerically obtained values. Error bars depict the standard
deviation for 50 different realizations of the noise potential
$V_{\varepsilon}(x)$. (b) The fidelity $\langle F(t)
\rangle_{noise}$ (circles), fidelity product $\langle F_P(t)
\rangle_{noise}$ (black dot-dashed lines), the values obtained via
$\det (\mathbf{PP}^\dag)$, where $P_{ij}=\sum_n a_n^{i*} a_n^j
\exp(i \omega_n t)$ is obtained via approximation presented by Peres
\cite{Peres} (blue line), and the fidelity obtained via trace-log
formula Eq. (\ref{trlogf}) (red dotted line). } 
\label{Gaussian}
\end{figure}

In Fig.~\ref{Gaussian}(a) we show the fidelity $\langle F(t)
\rangle_{noise}$ as a function of time, for three different numbers
of particles, $N=10$, $20$ and $50$ with $\varepsilon=0.05$. We find
that in the TG regime, the Loschmidt echo decays as a Gaussian:
$\langle F(t) \rangle_{noise}=\exp(-\langle
\lambda(N,\varepsilon)\rangle t^2)$; solid black lines represent the
Gaussian curves fitted to the numerically obtained values. We point
out that in every single realization of the noise, the fidelity for
a TG gas decays as a Gaussian, with small fluctuations in the value
of the exponent. The error bars in Fig.~\ref{Gaussian}(a) represent
the standard deviation of the fidelity at a given time. Note that
the standard deviation gets smaller with increasing particle number
$N$, which means that for sufficiently large $N$ it suffices to
calculate fidelity decay for a single realization of the potential
to obtain reliable values for $\lambda(N,\varepsilon)$. It s
interesting to compare the fidelity with the fidelity-product
$\langle F_P(t) \rangle_{noise}$, which can also be fitted well with
the Gaussian function as illustrated in Fig.~\ref{Gaussian}(b). We
find that the product $\langle F_P(t) \rangle_{noise}$ is
systematically below the value of the fidelity. 

\begin{figure}[!h]
\centering
\includegraphics[scale=0.48]{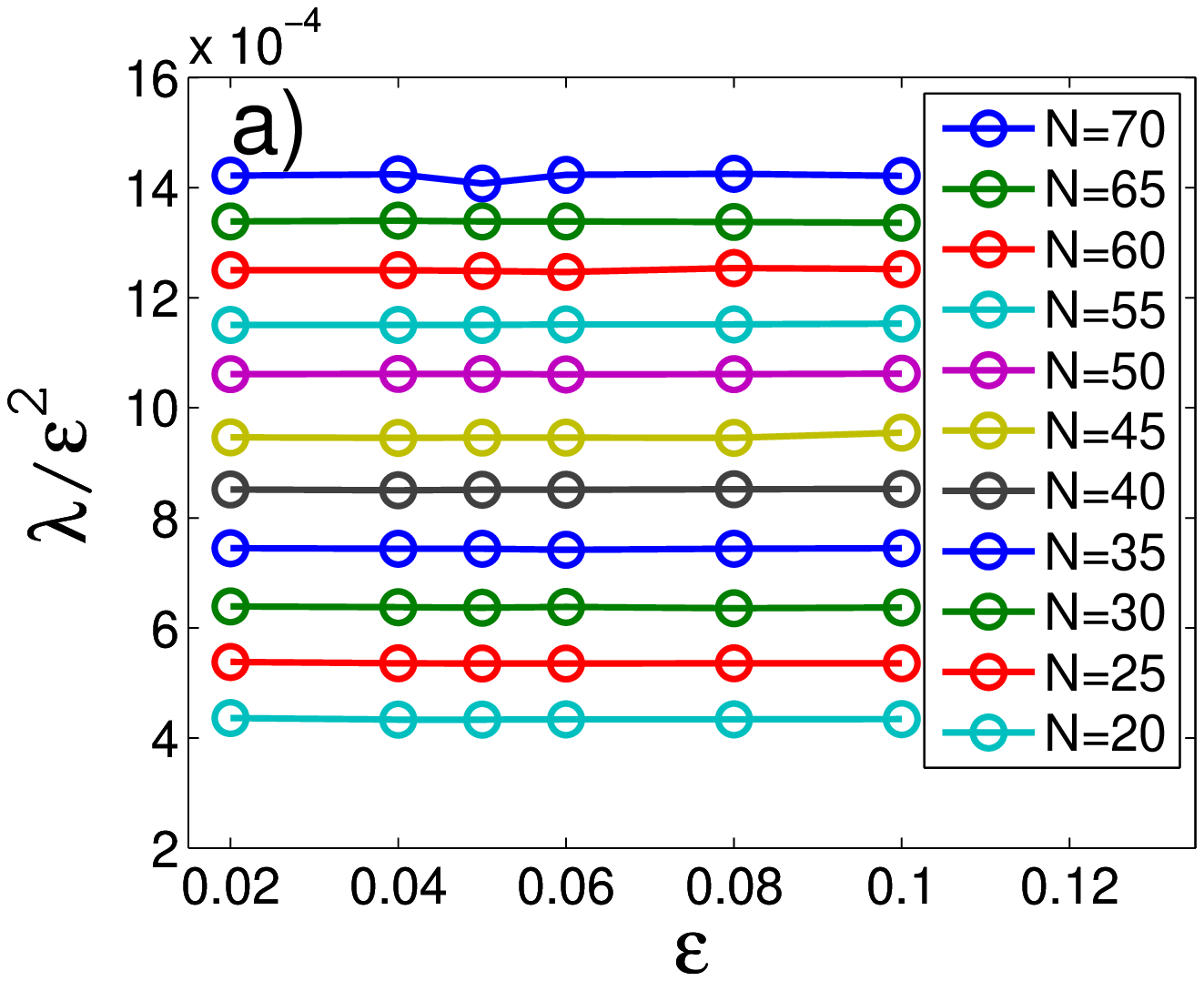}
\includegraphics[scale=0.48]{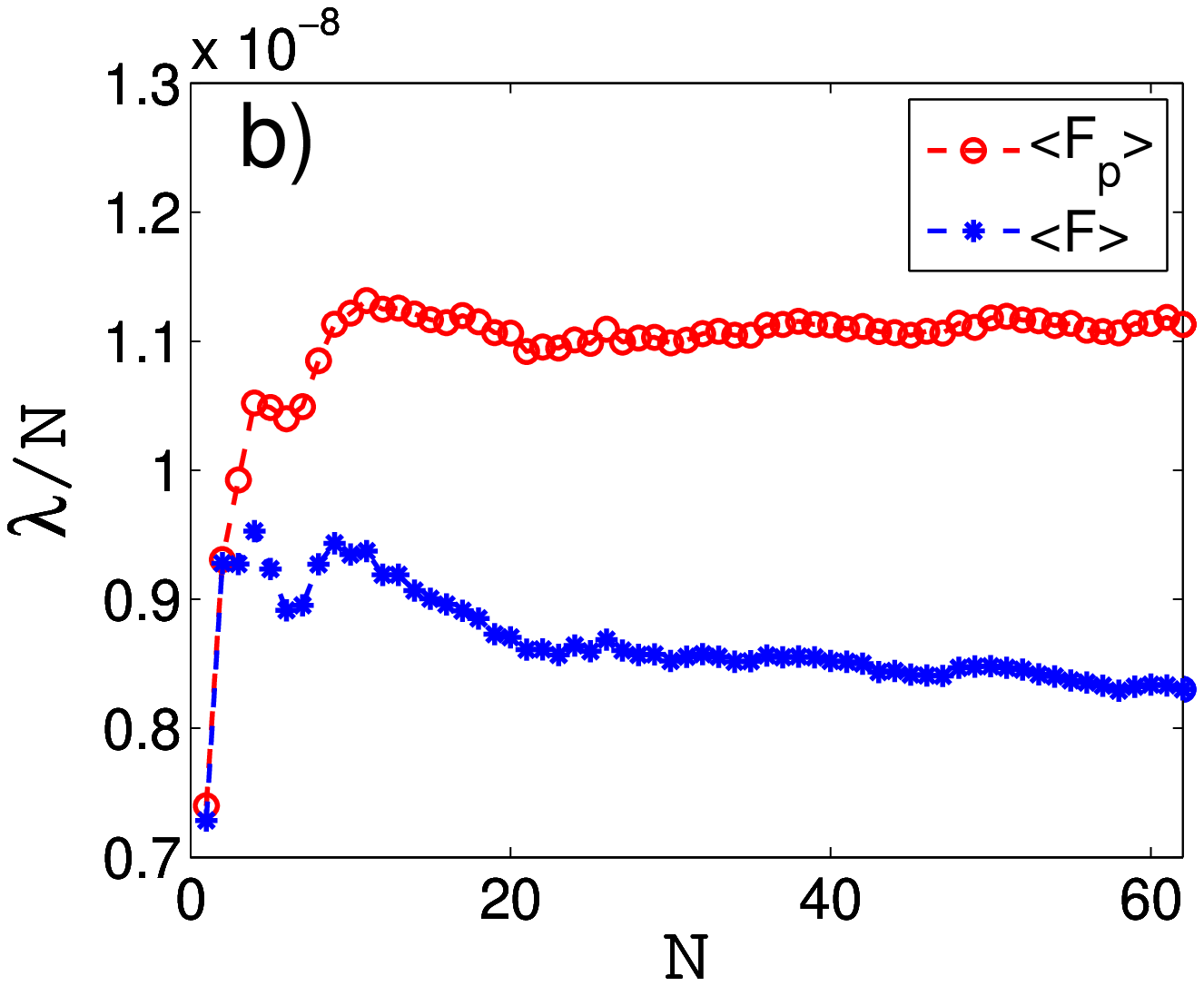}
\caption{(color online) The Gaussian exponent of the fidelity as a
function of $\varepsilon$ and $N$. 
(a) The quantities $\langle
\lambda \rangle/\varepsilon^2$ are plotted for different particle
numbers, and they are ordered just as in the legend (higher lines
are for larger values of $N$); obviously $\langle \lambda
\rangle\propto \varepsilon^2$. 
(b) The quantities $\langle \lambda_P
\rangle/N$ (red circles, upper line) and $\langle \lambda \rangle/N$
(blue asterisks, lower line), are plotted as a function of $N$ for $\varepsilon=0.05$. For
larger $N$ the lines become horizontal indicating that $\langle
\lambda \rangle\propto N \propto \langle \lambda_P \rangle$. }
\label{lambda}
\end{figure}

In Fig.~\ref{lambda} we depict the dependence of the fidelity, that
is, of the exponent $\langle\lambda(N,\varepsilon)\rangle$, on the
number of particles $N$ and $\varepsilon$. In Fig.~\ref{lambda}(a)
we plot $\langle\lambda(N,\varepsilon)\rangle/\varepsilon^2$ 
as a function of $\varepsilon$ for
different values of $N$; evidently we have
$\langle\lambda(N,\varepsilon)\rangle\propto \varepsilon^2$. In
Fig.~\ref{lambda}(b) we plot
$\langle\lambda(N,\varepsilon)\rangle/N$ as a function of $N$ for
$\varepsilon=0.05$; we clearly see that
$\langle\lambda(N,\varepsilon)\rangle\propto N$ for sufficiently
large $N$ (already for $N>20$).

In order to understand numerical results of Fig.~\ref{Gaussian} and 
Fig.~\ref{lambda}, we analytically explore the properties of the 
fidelity for a single realization of $V_{\varepsilon}(x)$. To this 
end we use an approximation from Peres \cite{Peres}, where to first 
order in $\varepsilon$ one has $\int dx \phi_j^{'*}\phi_i \approx 
\delta_{ij}$, and $a_{m}^{j}\approx a_{m}^{j'}$. The elements of the 
matrix $\mathbf{P}$, which yield the fidelity via Eq. (\ref{fidelityTG}), 
are then written as $P_{ij}=\sum_n a_n^{i*} 
a_n^j \exp(i \omega_n t)$, where $\omega_n=E_n'-E_n\approx \langle 
\phi_n|V_{\varepsilon}|\phi_n\rangle $. In Fig.~\ref{Gaussian}(b) we 
plot the fidelity obtained with this approximation (blue line) and 
the one obtained with the exact numerical evolution (blue circles); the 
agreement is excellent. The diagonal elements $|P_{ii}(t)|^2$ can be 
interpreted as single-particle fidelities corresponding to the 
initial states $\psi_i(x,0)$. It is straightforward to see that 
$|P_{ii}(t)|^2=\sum_{n,m} |a_n^{i}|^2 |a_m^i|^2 
\cos[(\omega_n-\omega_m) t]$, however, we note that in our 
simulations only a few terms contribute to the sum above, yielding 
oscillatory behavior of the single particle fidelities with 
relatively high amplitudes of the oscillation. 
Now we turn to the TG gas and our observation that the decay of 
fidelity is Gaussian. In order to derive this we use the 
trace-log formula for the determinants:
\begin{equation}
F(t)=\exp(\textrm{Tr}(\log (\mathbf{PP}^\dag))).
\end{equation}
We can approximate $\mathbf{PP}^\dag\approx 
\mathbf{1}+\mathbf{Q_1}t-\mathbf{Q_2}t^2+\mathcal{O}(t^3)$, 
where $\Delta_{nm}=\omega_n-\omega_m$,
$[Q_1]_{ij}=i\sum_{k=1}^N\sum_{n,m} a_n^{i*}a_n^{k}a_m^{j}a_m^{k*} 
\Delta_{nm}$, and $[Q_2]_{ij}=\frac{1}{2}\sum_{k=1}^N\sum_{n,m} 
a_n^{i*}a_n^{k}a_m^{j}a_m^{k*}\Delta_{nm}^2$. Next we expand the 
logarithm in trace-log formula which yields
\begin{equation}
F(t)=\exp(-\textrm{Tr}\mathbf{Q_2}t^2),
\label{trlogf}
\end{equation}
i.e., a Gaussian function. 
In our derivation we used $\textrm{Tr}\mathbf{Q_1}=0$. 
Red dotted line in Fig. \ref{Gaussian}(b) shows that Eq. (\ref{trlogf}) is 
an excellent approximation for larger $N$. 
The dependence of $\langle \lambda(N,\varepsilon)\rangle$ on $\varepsilon$ 
follows from the fact that $\Delta_{nm}^2 \propto \varepsilon^2$, 
whereas $\textrm{Tr}\mathbf{Q_2}\propto N$ (see Fig. \ref{lambda}).

In the rest of this section we argue that $F_P(t)<F(t)$, i.e., that 
the fidelity product is smaller than the fidelity. 
Obviously we need only diagonal elements of matrix $\mathbf{PP}^\dag$ 
to construct either $F(t)$ or $F_P(t)$, which we write as 
\begin{equation}
(\mathbf{PP}^\dag)_{ii}=|P_{ii}(t)|^2+
\sum_{k=1,k\neq i}^N|P_{ik}(t)|^2. 
\label{ppkriz}
\end{equation}
For the first term we can write $|P_{ii}(t)|^2=1-\alpha_i(t)$ where
$\alpha_i(t)$ is some function of time with properties
$\alpha_i(0)=0$ and $0\leq\alpha_i(t)\leq 1$ due to relation
(\ref{matrixP}). For the second term we write $\sum_{k=1,k\neq
i}^{N}|P_{ik}(t)|^2=\beta_i(t)$ where $\beta_i(0)=0$ and
$\beta_i(t)\geq0$. It follows that $(\mathbf{PP}^\dag)_{ii}=1-(\alpha_i(t)-\beta_i(t))$. 
By applying the trace-log formula in the same manner as before, we get for the fidelity
$F(t)=\exp(-\sum_{i=1}^N\alpha_i(t)+\sum_{i=1}^N\beta_i(t))$. 
The fidelity product corresponds to $\exp(-\sum_{i=1}^N\alpha_i(t))$, 
which yields $F(t)=F_P(t)\exp(\sum_{i=1}^N\beta_i(t))$; since
$\sum_{i=1}^N\beta_i(t)\geq0$ we have $F(t)\geq F_P(t)$.
Averaging over noise does not change this relation. 

\section{Fidelity in the mean field regime}

In this section we consider Loschmidt echo in the mean field regime,
that is, by employing the Gross-Pitaevskii theory. The
dynamics of Bose-Einsten condensates (BECs) is within the framework
of this theory described by using the nonlinear Schr\" odinger
equation (NLSE), which we write in dimensionless form:
\begin{equation}
i\frac{\partial \Phi(x,t)}{\partial
t}=\left[-\frac{\partial^2}{\partial
x^2}+V(x)\right]\Phi(x,t)+\tilde{g}_{1D}N|\Phi(x,t)|^2\Phi(x,t),
\label{GP}
\end{equation}
where $\tilde{g}_{1D}=2m X_0 g_{1D}/\hbar^2$ is the dimensionless
coupling strength and $\int|\Phi(x,t)|^2dx=1$; here we choose
$\tilde{g}_{1D}=0.04$, which can be experimentally obtained by
tuning the transverse confinement frequency to
$\omega_{\bot}/2\pi\approx 240$~Hz, and with $N=50$. With those
parameters the system is in the mean field regime with
$\gamma\approx 0.01$. All parameters are identical as in the
simulations of a TG gas, except that now $\omega_{\bot}$ is smaller.

To compute the fidelity of interacting BECs we repeat the same
procedure as for the TG gas: first we prepare the condensate in the
ground state of the container like potential $V_{L}(x)$ (i.e. we
solve numerically the stationary NLSE), second we suddenly expand
the container to $V_{2L}(x)$, and solve numerically the
time-dependent NLSE in the expanded potential without noise
[$V_{2L}(x)$], and with noise [$V_{2L}'(x)$], with identical initial
conditions. This gives us $\Phi(x,t)$ and $\Phi'(x,t)$ from which we
calculate the fidelity
\begin{equation}
F_{GP}(t)=|\int\Phi^{'*}(x,t)\Phi(x,t)dx|^2. 
\end{equation}
However, note that since we investigate the fidelity of a gas with 
$N$ particles, the mean-field $N$-particle wavefunction is 
a product state, $\psi_{GP}(x_1,\ldots,x_N,t)=\prod_{j=1}^{N} \Phi(x_j,t)$, 
and therefore the $N$-particle mean-field fidelity is 
\begin{equation}
F_{GP}^{N}(t)=|\int\psi_{GP}^{'*} \psi_{GP} dx_1\ldots dx_N|^2=[F_{GP}(t)]^{N}. 
\label{FGPN}
\end{equation}
Finally, we average over 50 different realizations of the potential to obtain 
$\langle F_{GP}(t)\rangle_{noise}$ and $\langle F_{GP}^{N}(t)\rangle_{noise}$.

\begin{figure}[!h]
\centering
\includegraphics[scale=0.45]{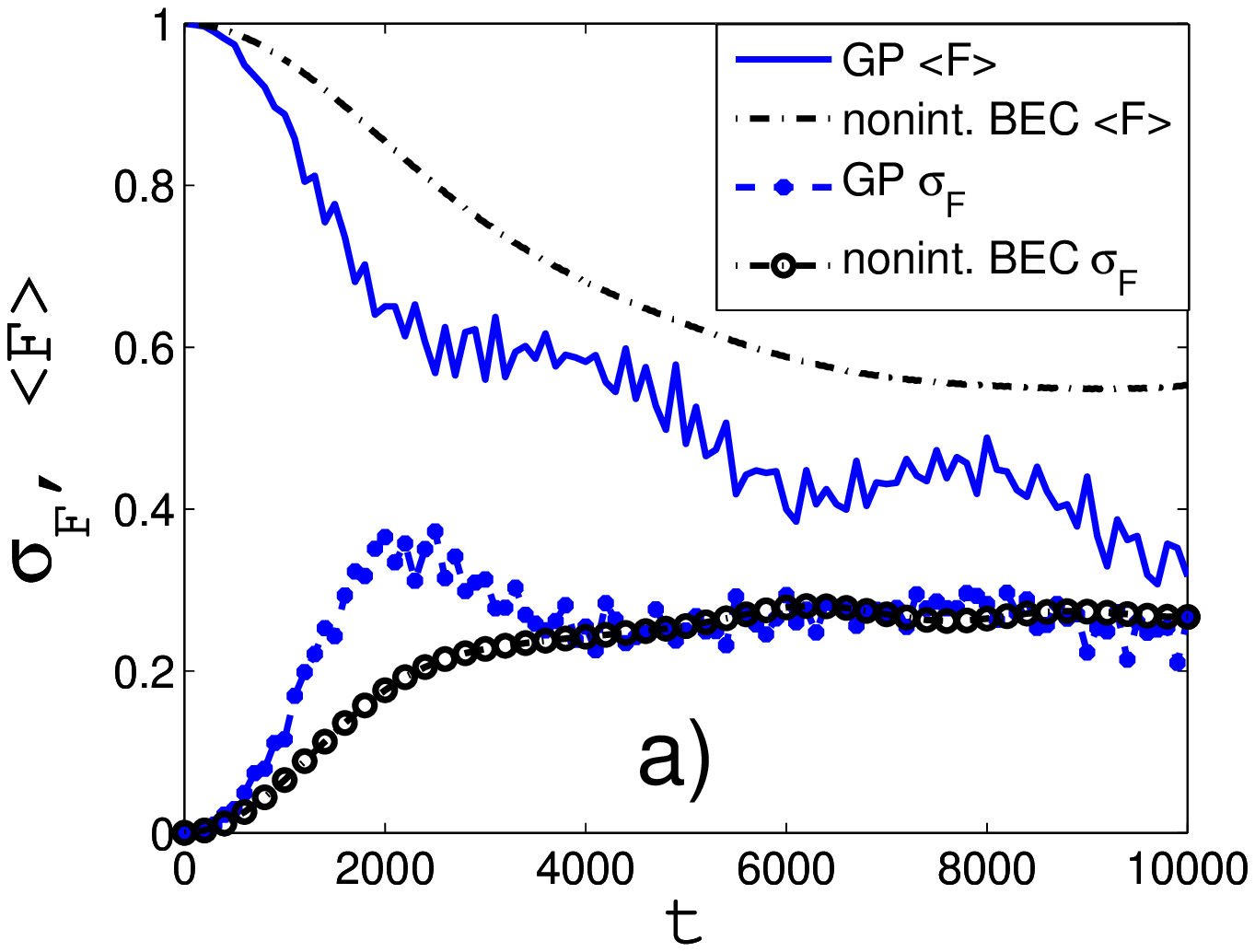}
\includegraphics[scale=0.45]{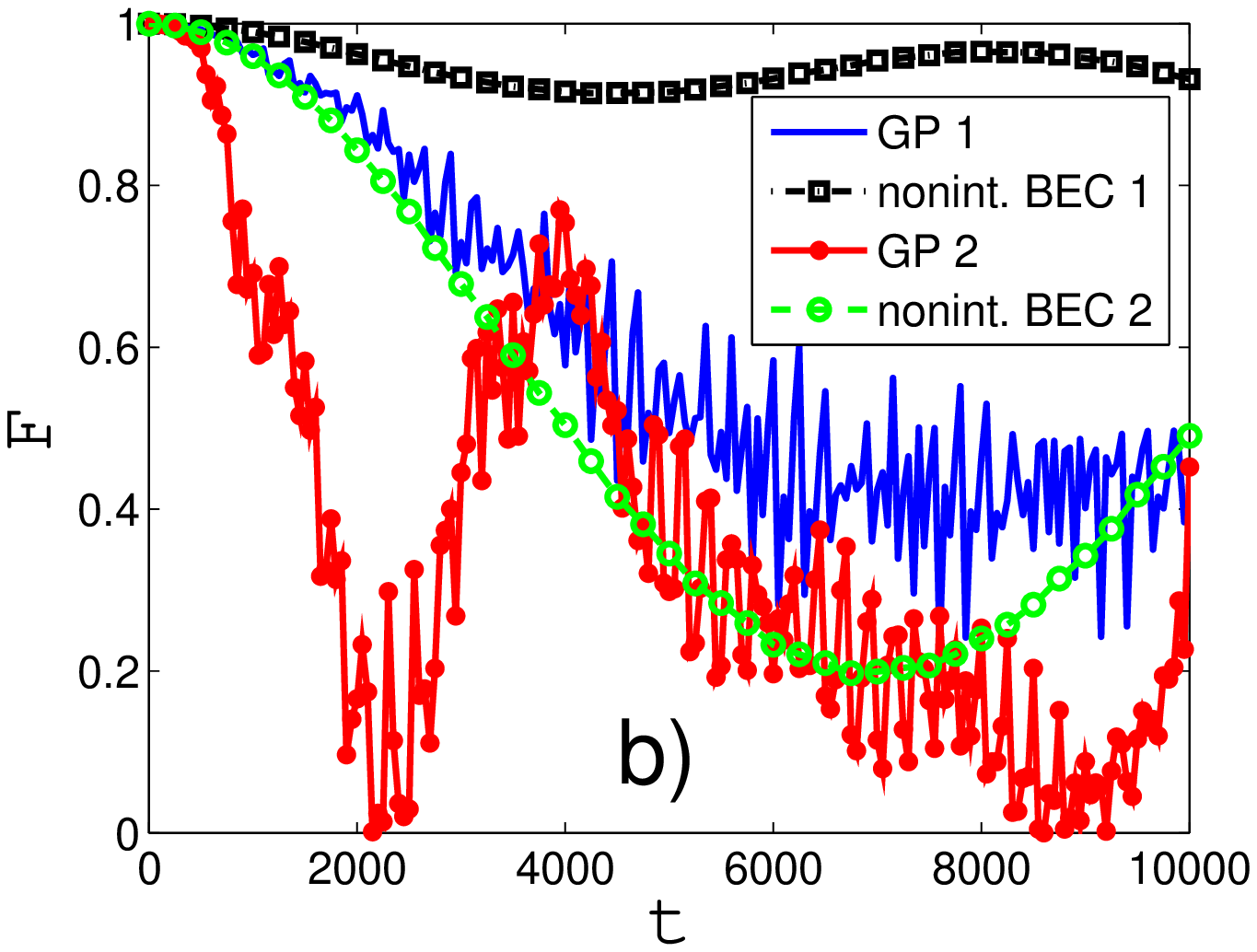}
\caption{(a) Noise-averaged fidelities of evolving BECs for $N=50$ and $\varepsilon=0.05$, 
and standard deviations from the noise-average. 
The averaged fidelity for a noninteracting BEC is shown with the black dot-dashed line, 
and its standard deviation with open black circles. 
The averaged fidelity for a weakly-interacting BEC is shown with the blue solid line, 
and its standard deviation with closed blue circles. 
(b) Fidelities of the evolving BECs for two different realizations of the 
noise potential. 
} 
\label{figGP}
\end{figure}

In Fig.~\ref{figGP}(a) we plot $\langle F_{GP}(t)\rangle_{noise}$ and its standard 
deviation for noninteracting and weakly-interacting BECs. We see that oscillations are 
superimposed on the overall decay in contrast to the TG gas case. 
We find that in the mean-field regime described by the Gross-Pitaevskii equation
the fidelity decays faster for larger nonlinearity (interaction strength). 
It is worthy to point out that $F_{GP}(t)$ is very dependent
on the particular realization of $V_{\varepsilon}(x)$, which is not
the case for the TG gas. This is illustrated in Fig.~\ref{figGP}(b) where we 
show dynamics of $F_{GP}(t)$ for two different realizations of the noise potential;
we observe a large dependence of $F_{GP}(t)$ on a particular realization of the noise. 
This is a consequence of the fact that the oscillation frequency of fidelity
$|P_{11}(t)|^2=\sum_{n,m} |a_n^{1}|^2 |a_m^1|^2
\cos[(\omega_n-\omega_m) t]$ for the noninteracting BEC essentially
depends on the difference between only several frequencies which is
very noise sensitive, and this behavior is inherited in the
nonlinear mean-field regime.

\begin{figure}[!h]
\centering
\includegraphics[scale=0.45]{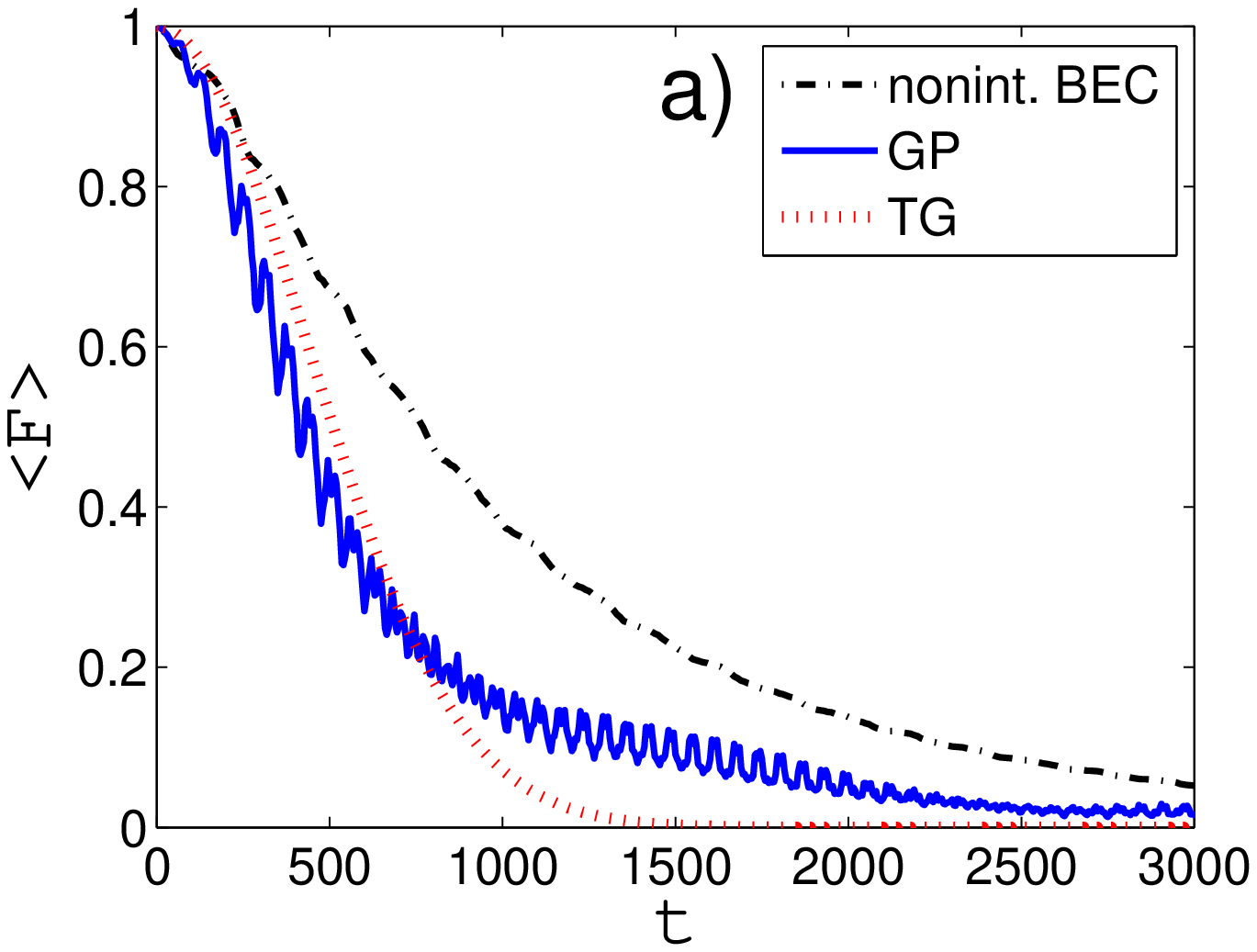}
\includegraphics[scale=0.45]{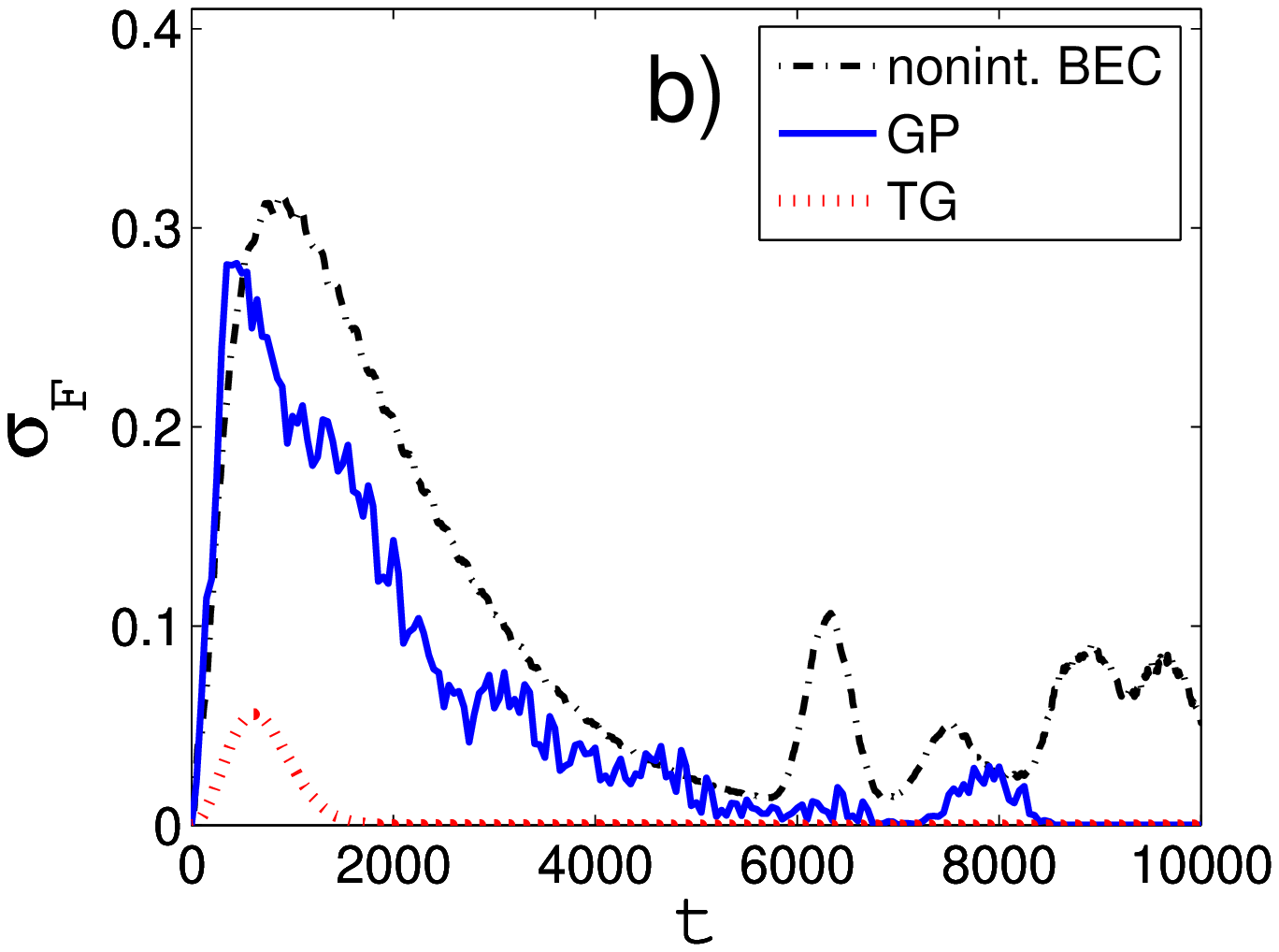}
\caption{Comparison of the averaged fidelities (a) and their standard deviations (b) 
for the TG gas ($\langle F(t)\rangle_{noise}$, red dotted line), the weakly-interacting BEC 
($\langle F_{GP}^{N}(t)\rangle_{noise}$, solid blue line), and the noninteracting BEC (black dot-dashed line), 
for the same number of particles. The parameters are $N=50$ and $\varepsilon=0.05$.} 
\label{comparison}
\end{figure}

Finally, in Fig. \ref{comparison} we compare the fidelities of the 
noninteracting BEC, the weakly-interacting BEC, and the TG gas. 
Note that for proper comparison one should compare $F_{GP}^{N}(t)$ with $F(t)$. 
We see that the mean-field fidelity shows richer behavior. 
In the regime of parameters we used, we find that $\langle F_{GP}^{N}(t)\rangle_{noise}$
decays faster than the TG regime fidelity in the first part of the 
decay dynamics, but later the mean-field regime fidelity decay slows down 
in comparison to the TG gas decay.

\section{Conclusion}

In conclusion, we have explored Loschmidt echo (fidelity) in two 
regimes of one-dimensional interacting Bose gases: the strongly 
interacting TG regime, and the weakly-interacting mean-field regime 
described within the Gross-Pitaevskii theory. 
The gas is initially in the ground state of a trapping potential that 
is suddenly broadened, and the decay of fidelity is studied numerically 
by using a small spatial noise perturbation. 
We find (numerically and analytically) that the fidelity of the TG gas 
decays as a Gaussian with the exponent proportional to the number of particles 
and the magnitude of the small perturbation squared 
(see Fig.~\ref{Gaussian} and Fig.~\ref{lambda}). 
Our results do not depend on the details of trapping potential; we have obtained 
the same behavior for a gas that is initially loaded in the ground state of the 
harmonic oscillator potential, which is subsequently suddenly broadened. 
Furthermore we find that Gaussian decay remains if we initiate the dynamics 
from some excited initial state or from a superposition of such states. 
In the mean-field regime the Loschmidt echo decays faster for 
larger interparticle interactions (nonlinearity), and it shows richer 
behavior than TG Loschmidt echo dynamics with oscillations superimposed on 
the overall decay (see Fig.~\ref{figGP} and Fig.~\ref{comparison});
it also has much larger sensitivity on the noise (see Fig.~\ref{figGP}(b)). 
Finally, we would like to mention that perhaps the most interesting regime of 
Loschmidt echo dynamics would be for intermediate Lieb-Liniger interactions, which 
seem to be exactly solvable only for specific external potential configurations
\cite{Jukic}. 

\acknowledgments

This work is supported by the Croatian Ministry of Science (Grant
No. 119-0000000-1015). H.B. acknowledge support from the
Croatian-Israeli project cooperation and the Croatian National
Foundation for Science. We are grateful to T. Gasenzer, R. Pezer 
for most useful discussions, and to J. Goold for pointing to us 
Ref. \cite{Goold2011}. 
The first investigations of the Loschmidt echo in 
TG gases were made in a Diploma thesis by Igor \v{S}egota,
at the Department of Physics, University of Zagreb, under supervision 
of H.B. 

{\it Note added.} In the first version (v1) of this work posted on the arXiv 
we have erroneously compared $F_{GP}(t)$ from the mean-field regime, with the 
fidelity $F(t)$ describing the TG regime. We emphasize that the true 
mean-field fidelity depends on the number of particles $N$ and one should compare 
$F_{GP}^{N}(t)$ [see Eq. (\ref{FGPN})] from the mean-field regime, with $F(t)$ as we did in our 
new Figure \ref{comparison}.

\end{document}